\documentclass[acmlarge,screen]{acmart}

\usepackage{adjustbox}
\usepackage{multirow}
\usepackage{booktabs,subcaption,dcolumn}
\usepackage{tikz}
\usepackage{pifont}
\newcommand{\cmark}{\ding{51}}
\newcommand{\xmark}{\ding{55}}
\usepackage{soul}

\newcolumntype{?}{!{\vrule width 1.5pt}}

\newcommand{\NA}{---}

\newcommand\pdate[1]{{\fontfamily{qcr}\selectfont #1}}
\newcommand\power[1]{{\fontfamily{qhv}\selectfont #1}}
\newcommand\dataset[1]{{\footnotesize{ #1}}}

\newcommand\revision[1]{%
  \bgroup
  \hskip0pt\color{blue!80!black}%
  #1%
  \egroup
}

\setcopyright{acmcopyright}
\acmJournal{DTRAP}
\acmYear{2021} \acmVolume{1} \acmNumber{1} \acmArticle{1} \acmMonth{1} \acmPrice{15.00}\acmDOI{10.1145/3469659}

\AtBeginDocument{%
  \providecommand\BibTeX{{%
    \normalfont B\kern-0.5em{\scshape i\kern-0.25em b}\kern-0.8em\TeX}}}

\begin{document}
\title{Modeling Realistic Adversarial Attacks against Network Intrusion Detection Systems}

\author{Giovanni Apruzzese}
\orcid{0000-0002-6890-9611}
\email{giovanni.apruzzese@uni.li}
\affiliation{%
  \institution{Institute of Information Systems, University of Liechtenstein}
  \city{Vaduz}
  \state{Liechtenstein}
}

\author{Mauro Andreolini} 
\email{mauro.andreolini@unimore.it}
\author{Luca Ferretti} 
\email{luca.ferretti@unimore.it}
\affiliation{%
  \institution{Department of Physics, Informatics and Mathematics, University of Modena and Reggio Emilia}
  \city{Modena}
  \state{Italy}
}

\author{Mirco Marchetti} 
\email{mirco.marchetti@unimore.it}
\affiliation{%
  \institution{Department of Engineering "Enzo Ferrari", University of Modena and Reggio Emilia}
  \city{Modena}
  \state{Italy}
}
\author{Michele Colajanni} 
\email{michele.colajanni@unimore.it}
\affiliation{%
  \institution{Department of Informatics, Science and Engineering, University of Bologna}
  \city{Bologna}
  \state{Italy}
}

\begin{abstract}
The incremental diffusion of machine learning algorithms in supporting cybersecurity is creating novel defensive opportunities but also new types of risks. Multiple researches have shown that machine learning methods are vulnerable to adversarial attacks that create tiny perturbations aimed at decreasing the effectiveness of detecting threats. We observe that existing literature assumes threat models that are inappropriate for realistic cybersecurity scenarios because they consider opponents with complete knowledge about the cyber detector or that can freely interact with the target systems. By focusing on Network Intrusion Detection Systems based on machine learning, we identify and model the real capabilities and circumstances required by attackers to carry out feasible and successful adversarial attacks.
We then apply our model to several adversarial attacks proposed in literature and highlight the limits and merits that can result in actual adversarial attacks. The contributions of this paper can help hardening defensive systems by letting cyber defenders address the most critical and real issues, and can benefit researchers by allowing them to devise novel forms of adversarial attacks based on realistic threat models.
\end{abstract}



\maketitle

\section{Introduction}
\label{sec:1-intro}

Machine Learning (ML) is increasingly adopted in multiple domains. As a typical consequence of this world, it is becoming a preferred target of modern adversarial attacks~\cite{Papernot:SoK, tabassi2019taxonomy}. This emerging digital threat involves attackers that exploit the intrinsic vulnerabilities of machine learning methods by leveraging specific inputs that thwart their predictions. The state of the art focuses on computer vision and natural language processing applications~\cite{Liu:Survey}. However, several studies on cybersecurity are appearing~\cite{Martins:Adversarial} especially because this scenario is inherently affected by opponents that are rarer in other considered contexts. In cybersecurity, the main interest has been on malware and spam detectors, while fewer works consider Network Intrusion Detection Systems (NIDS) that are of interest for this paper~\cite{DeLucia:Adversarial,Apruzzese:Addressing, Pitropakis:Taxonomy}.

The motivations for our research starts from the observation that most papers proposing, evaluating and considering NIDS in adversarial scenarios do not discuss the realistic level of the proposed attacks. A typical research work may assume any threat model, and then proceed to analyze the effects of the attack with no or insufficient considerations about the feasibility of the considered scenario~\cite{Ibitoye:Threat}. For example, some papers assume attackers that know everything about the target system~\cite{Hashemi:Towards}. Others suppose that an adversary can perform an arbitrarily large number of trials against the NIDS without getting noticed~\cite{Peng:Adversarial}.
Although investigating the effectiveness of adversarial attacks against any ML system is an important goal for creating more robust detectors, cybersecurity scenarios should always deal with realistic issues and adversaries. Otherwise, defenders may spend resources against false hits or unrealistic problems, when more critical issues are taking place.
The sheer amount of researches on adversarial attacks may even induce cyber defenders to think that any Machine Learning-based NIDS (ML-NIDS) is an unreliable defensive system, although this is not the case.

This paper analyzes the realistic feasibility of adversarial attacks against ML-NIDS by identifying the capabilities and conditions that are necessary for carrying out such attacks against ML-NIDS. We identify five elements of the target system that can be leveraged to perform adversarial attacks. By introducing the concept of \emph{power}, which models the attacker's knowledge and capabilities on each of these five elements, we outline the realistic circumstances that allow an attacker to thwart the target system. Our proposal complements taxonomies that do not delve into the feasibility and constraints of real ML-NIDS scenarios (e.g.,~\cite{Huang:Adversarial, Laskov:Practical}) and it can be applied to assess and evaluate adversarial attacks against any ML-NIDS.

We apply the proposed concepts to analyze the state of the art on adversarial attacks against ML-NIDS. We assess the feasibility levels of current researches, and analyze the details of four significant use cases. We can conclude that few papers assume scenarios that are representative of an authentic cyber defensive world. Hence there is a gap between academic and real environments that must be filled for a mutual interest. The most recent efforts are taking into account more interesting and realistic scenarios, which is a positive trend.

Researchers on adversarial ML can benefit of this paper by formulating original adversarial attacks that assume realistic threat models. Moreover, security experts working on ML-NIDS can harden their defensive systems by considering the smaller and more realistic subset of feasible attacker's scenarios.

The paper is structured as follows. 
Section~\ref{sec:background} presents the basic concepts of this paper.
Section~\ref{sec:related} motivates and compares our paper against related work.
Section~\ref{sec:modeling} proposes our original analysis for modeling realistic adversarial attacks against ML-NIDS.
Section~\ref{sec:analysis} applies the proposed analysis to existing proposals of adversarial attacks, and describes three case studies.
Section~\ref{sec:conclusions} concludes the paper with some final remarks and possible future work.
\section{Background}
\label{sec:background}

We consider adversarial attacks against network intrusion detection systems that rely on machine learning methods for detecting attacks. Let us summarize these concepts.

\subsection{Network Intrusion Detection Systems based on Machine Learning}
\label{sec:detection}

The detection of malicious events is a prominent issue in the cybersecurity landscape. As manual inspection is impossible when millions of events per day occur, human defenders are supported by intrusion detection systems (IDS) that analyze data from different sources and, when specific conditions are met, generate alerts for the triage phase~\cite{Liao:Review}.
We consider the specific category of Network-IDS, which aim at identifying intrusions at the network traffic level.
Several types of NIDS exists, but common differences involve the data type analyzed, and the method used to perform the detection.

The first generation of NIDS used to analyze network packets by inspecting their payload. While this approach may be more accurate, it cannot be applied when data is encrypted, and requires high amounts of computational resources to process each network packet. The exponential growth of traffic that often is encrypted raised the interest towards NIDS inspecting metadata, such as network flows~\cite{Liao:Review}. In the last decade~\cite{Sperotto:flowIDS} many NIDS leverage the analysis of network metrics that summarize entire communication sequences between two endpoints, such as the duration of the session or the amount of exchanged bytes. This information is computationally feasible to store and analyze, and does not present privacy concerns.

With regards to the detection method, initial NIDS identified known threats on the basis of human-written \textit{signatures}~\cite{Denning:Intrusion}, but recent solutions adopt (also) data-driven methods \cite{Marchetti:Countering} leveraging anomaly detection typically based on machine learning techniques~\cite{Gardiner:Malware, Sommer:Outside}. These approaches allow a NIDS to detect also unknown malicious events that expert attackers may adopt to evade signature-based methods. 

This paper focuses on ML-NIDS. ML generates models that are able to learn specific patterns by providing them with training data. These patterns are then used to make predictions on new and unseen sets of data~\cite{Papernot:SoK}. The main advantage of these detection schemes is their capability of automatically learning from training data without human intervention, thus simplifying the resource-intensive management procedures required by traditional misuse-based approaches. Furthermore, they are also able of detecting novel attack variants, for which no known signature exists and that would be undetectable by NIDS based exclusively on rules. There exists a wide literature on ML-NIDS~\cite{Gardiner:Malware, Buczak:Survey, Berman:Survey}, which can work either on payload (e.g.,~\cite{zanero2004unsupervised}) or on netflow data (e.g.~\cite{Bilge:Disclosure}). Proposals include supervised and unsupervised techniques, and also an increasing number of recent approaches based on deep neural networks~\cite{Martins:Adversarial}. Although the detection performance depends on the ML algorithm, the superiority of deep learning methods in this domain is still questionable~\cite{Apruzzese:Deep}.

\begin{figure}[!htbp]
    \centering
    \includegraphics[width=0.7\columnwidth]{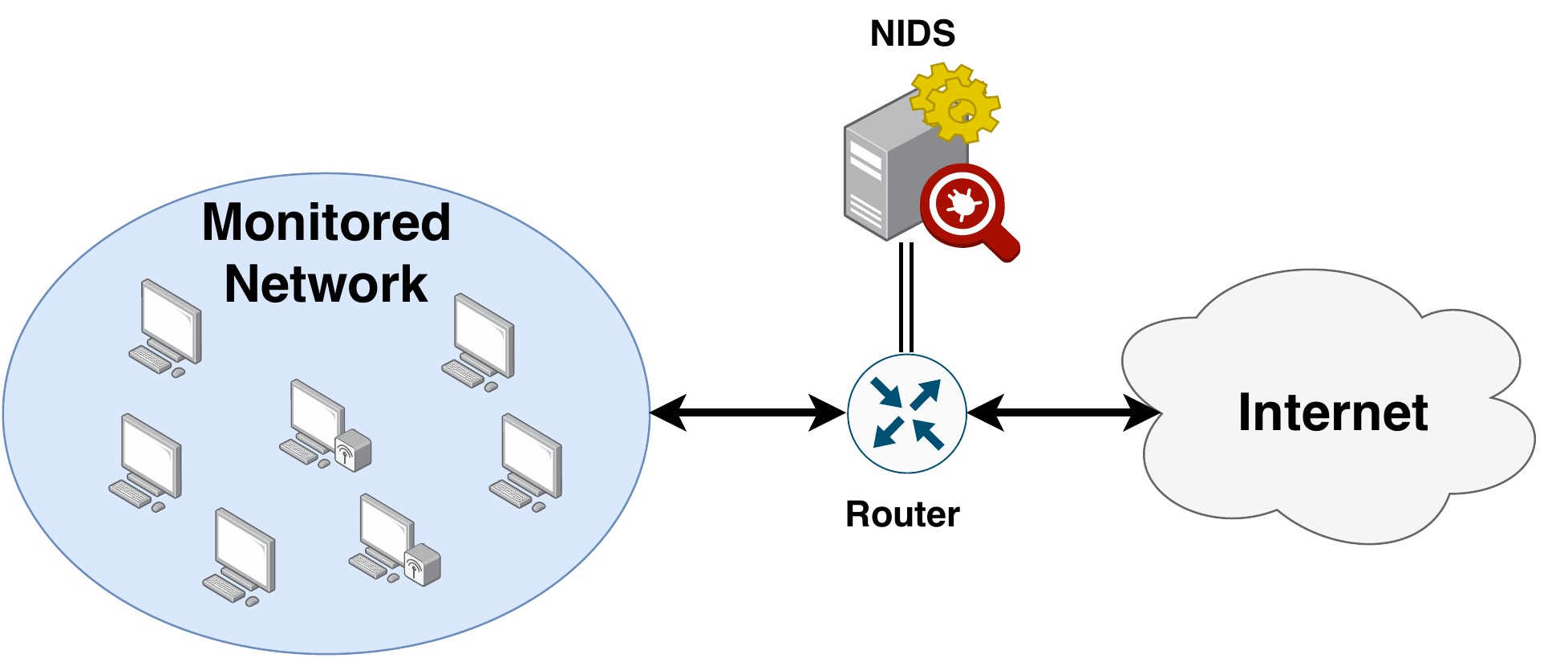}
    \caption{Typical deployment of a Network Intrusion Detection System.}
    \label{fig:scenario}
\end{figure}

A typical deployment scenario is represented in Figure~\ref{fig:scenario}, where the border router forwards all incoming traffic to the NIDS, which proceeds to analyze it. This system is one of the most protected elements of the entire network. Cybersecurity threat models exclude that an attacker can access the NIDS because, in a similar instance, the entire defensive infrastructure can be considered compromised: any cybersecurity measure can be easily bypassed if the attacker acquires direct or physical access to the defensive system~\cite{Shetty:Simulation}.

\begin{figure}[!htbp]
    \centering
    \includegraphics[width=0.7\columnwidth]{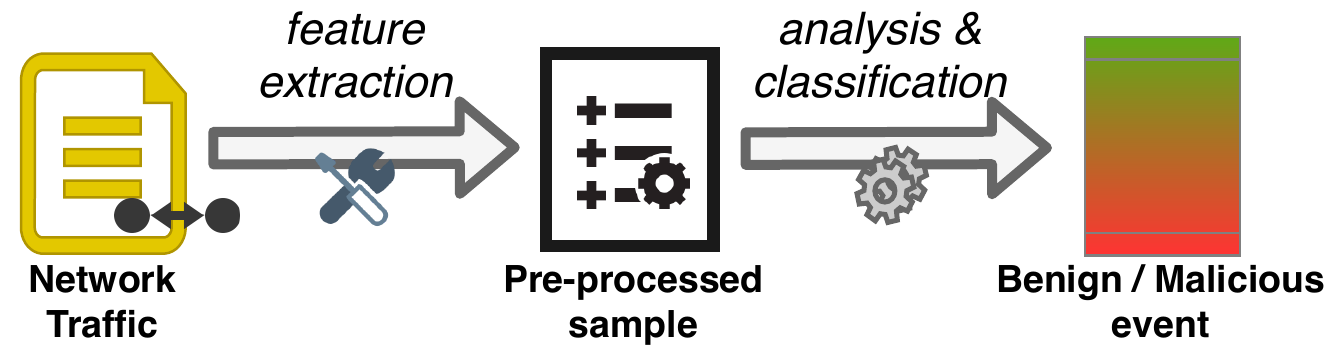}
    \caption{Workflow of a ML-NIDS.}
    \label{fig:ML-NIDS}
\end{figure}

In Figure~\ref{fig:ML-NIDS} we report the typical workflow of a ML-NIDS. The network traffic undergoes some pre-processing operations that extract the relevant features from the data, transforming it into samples accepted by the ML model. These samples are then forwarded to the (trained) ML model that will analyze them and determine whether they are legitimate or not.

There exist several solutions of ML-NIDS. An organization may adopt and maintain a ML-NIDS on premises, for example by leveraging open-source software~\cite{Shanthi:Detection}. Alternatively, it may rely on third-party products~\cite{Shetty:Simulation,Chernikova:Adversarial,Berman:Survey, Shenfield:Intelligent}.

In the former case, the life-cycle of the ML model is managed by the organization, which has access to all the ML components, including the model's internal configuration parameters, the feature set, and the dataset used for training purposes. This dataset can be built, for example, by using the regular network traffic generated in the monitored enterprise as ``benign'' samples, while traces of malicious traffic can be obtained either synthetically or through external security feeds~\cite{Garcia:CTU}. The dataset can also be periodically updated to better resemble the modifications of the network environment, or to include malicious samples of novel attack variants. The updated version will then be used to re-train the model accordingly~\cite{Li:Chronic}.

If the ML-NIDS belongs to a third party vendor, then the ML model is trained on datasets that are likely unavailable to the organization. Hence, all the training and management operations will be performed by the third-party vendor. In a similar scenario the organization has little or no information about the ML-model adopted by the NIDS. It is used as a black-box, where the network traffic generated by the organization is inspected by the model, whose results are then provided to the organization, usually in the form of alerts. An organization may be able to add some rules, for example to denote the most relevant server machines, but the internal configuration and the parameters of the ML model are unknown and not observable~\cite{Shetty:Simulation,Chernikova:Adversarial}.

ML-NIDS are gaining popularity~\cite{Gardiner:Malware, Apruzzese:Deep}, with the typical consequence that real attackers are turning their attention to the vulnerabilities of some ML components. This paper considers the so called adversarial attacks against ML-NIDS~\cite{Martins:Adversarial, Truong:Artificial}, and does not make any assumption on the specific ML-algorithm, nor on the analyzed data type of the ML-NIDS.

\subsection{Adversarial Attacks against Machine Learning}
\label{sec:attacks}

Adversarial attacks involve the application of perturbations to some data with the goal of fooling the ML model, thus resulting in an incorrect detection output that favors the attacker. These perturbations should be imperceptible to a human observer. A similar objective is easily verifiable in computer vision because the modification of few pixels causes unexpected results (e.g.,~\cite{Su:One}). It is more difficult to reveal perturbations when the manipulation is performed on network traffic data because each domain has its own characteristics. Attacks that are effective in a scenario may be unsuccessful in others~\cite{Hashemi:Towards, Sharif:Accessorize}.

An adversarial attack can affect the training phase of the machine learning algorithm: in this case, it is denoted as \textit{poisoning} attack, where the aim is to influence the detection through manipulations of the training dataset. Manipulations may involve either the injection of new data samples or the modification of existing samples (e.g., by flipping the labels of some data points~\cite{Kloft:Online}).

Alternative attacks may occur at \textit{inference}-time when the model is already operational. The focus is to thwart detection by leveraging the sensitivity of the (trained) model to its learned decision boundaries.
The most renown adversarial attacks in cyber detection are denoted as \textit{evasion} attacks, where the goal is misclassifying malicious samples as legitimate~\cite{Laskov:Practical}.

Literature presents three main cases of adversarial attacks, depending on the characteristics of the threat model.
The first case involves an opponent that knows everything about the target system. By leveraging such information it is possible to understand the decision boundaries of the ML model and to craft specific samples that thwart the detection mechanism. These attacks are also known as \textit{white-box} attacks~\cite{Martins:Adversarial}.

The second type of instances, denoted as \textit{black-box} adversarial attacks, assumes that the attacker has no information about the detection mechanism, but they are able to query the machine learning model by issuing some samples and inspecting the resulting classification. The attacker can exploit this input/output association to acquire information about the model adopted for detection. Such information can be leveraged in two ways.

\begin{itemize}
\item
The attacker can repeatedly modify the malicious samples until they are misclassified. The goal is to determine the boundaries used by the model to distinguish benign from malicious samples.
\item Alternatively, the attacker can create a surrogate model of the detection system, and then foster the \textit{transferability} property of ML to devise adversarial samples that fool the surrogate classifier and, in turn, also the real detector~\cite{Papernot:Practical}. 
\end{itemize}
Literature refers to ML models that are exploited through similar attacking strategies as ``oracles''~\cite{Papernot:SoK}.

In the third case, denoted as \textit{gray-box} attack, the adversary is more constrained and has limited knowledge about the classifier~\cite{Huang:Malware}. 
For example, they may only know a subset of the features adopted by the ML model; or they may know which algorithm is being used, but without any information on its learned configuration settings.

Some ML algorithms can be more resilient than others. For instance, gradient-based attacks (e.g.~\cite{Ayub:Model}) are more effective against neural networks than against tree-based methods. However, any ML algorithm is intrinsically vulnerable to adversarial attacks~\cite{ilyas2019adversarial}. Thus, in this paper we do not focus on any specific ML method, and our analyses and conclusions can be applied to any ML-NIDS.
\section{Related Work}
\label{sec:related}

Three trends motivate this paper: the increasing reliance of cyber defenses on machine learning and artificial intelligence tools, the consequential rise of novel threats based on adversarial attacks, and the lack of comprehensive guidelines detailing how to fight this emerging menace. We describe each of these points and compare our paper with related researches on adversarial machine learning in cybersecurity.

The adoption of techniques belonging to the machine learning paradigm (and to the broader artificial intelligence concept) is growing. Modern advanced systems require the execution of a large number of tasks that cannot be entirely managed by human personnel~\cite{andreolini2014monitoring}. Hence, the promising automation capabilities of artificial intelligence methods are greatly appreciated. As an example, autonomous agents are being developed for military applications~\cite{Kott:Doers}, as well as for the deployment of 5G services~\cite{Cayamcela:Artificial}, and also for cyber security~\cite{Xin:Machine,Apruzzese:Deep}.
The increasing diffusion of machine learning leads to questioning their efficacy and trustworthiness when deployed in critical environments~\cite{Guan:Machine, Varshney:Trustworthy, Buczak:Survey}, as it has been observed even by the European Commission~\cite{Europe:AI}. 

\begin{figure}[!htbp]
    \centering
    \includegraphics[width=0.4\columnwidth]{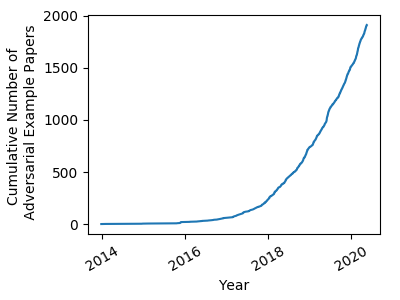}
    \caption{Growth of Adversarial Machine Learning papers over the years (source:~\cite{AdversarialML:URL}).}
    \label{fig:adv_papers}
\end{figure}

Among the issues affecting ML methods, the scientific community is giving importance to the effectiveness of these algorithms when they are subject to adversarial attacks~\cite{Papernot:SoK}. 
The amount of research papers on adversarial machine learning have rapidly increased as evidenced by Figure~\ref{fig:adv_papers} (taken from~\cite{AdversarialML:URL}).
In this broad landscape, the majority of studies are on adversarial attacks against image and text classification (e.g.,~\cite{Liu:Survey,Akhtar:Threat,Zhang:Adversarial2020,Zugner:Adversarial, Yuan:Adversarial}), and only a small percentage looks at the problem from a cybersecurity perspective, which is the focus of this paper. Moreover, most papers consider a multitude of scenarios where the target model suffers significant performance degradation (for example, a proficient model can be fooled with over 90\% confidence~\cite{Martins:Adversarial}) that may result in overestimating the actual problem posed by adversarial attacks in real environments. Our work aims to answer the question about which adversarial attacks proposed in literature are a real threat to modern cyber systems.

Let us compare our paper with related researches on adversarial attacks against NIDS. Among the first studies that summarize this threat we mention the seminal work in~\cite{Corona:Adversarial} which did not consider attacks against ML-NIDS. The authors in~\cite{Gardiner:Malware} evidence the problem of botnet detection in adversarial scenarios, but the considered attacks were oriented towards more general applications of machine learning (such as~\cite{Kantchelian:Evasion}), and they do not discuss any use-cases of adversarial perturbations against NIDS. 
Two significant surveys on adversarial machine learning were published by Biggio et al.~\cite{Biggio:Wild} and by Papernot et al.~\cite{Papernot:SoK}. These papers contain a comprehensive review of the state of the art, although they consider attacks in multiple domains and are do not address the intrinsic issues affecting NIDS scenarios. A similar difference characterizes also some recent reviews~\cite{Pitropakis:Taxonomy, Qiu:Review}. 
The exhaustive work of~\cite{Sadeghi:Taxonomy} proposes several taxonomies for adversarial attacks at different stages, but it does not delve into the characteristics of cyber detection contexts. Recent works addressing this threat from a cybersecurity perspective can be found in~\cite{DeLucia:Adversarial, Apruzzese:Addressing, Martins:Adversarial}, which summarize adversarial ML for cyber detection, but they do not consider the realistic level of the discussed papers. The review in~\cite{Ibitoye:Threat} proposes the original concept of adversarial risk that is used to measure the likelihood that an adversarial attack is successful, but this concept is based on the vulnerabilities of a machine learning approach, and does not take into account real world constraints, which is the focus of our contribution. Some practical applications of adversarial attacks not regarding cybersecurity are presented in~\cite{Papernot:Practical} and in~\cite{Zhou:Survey} which adopts a game-theoretical perspective. 
Unlike all these papers, our study provides an original modeling and analysis of adversarial attacks against ML-NIDS that focus on the realistic feasibility of the proposed threats. Our work is oriented to both ML researchers and security specialists that are interested in evaluating their cyber defence systems in real scenarios that involve constrained attackers.

The recent analysis of Kumar et al.~\cite{Kumar:Adversarial} presents the perspective of real enterprises about adversarial attacks and concludes that modern organizations are aware of these problems, but do not consider this threat as a top-priority because there are no defensive mechanisms that are truly effective in real environments. This conclusion evidences that most literature on adversarial attacks considers unfeasible scenarios. Hence we find useful to outline the main characteristics that must be considered to reproduce realistic adversarial settings, which can be used to devise sensible attacks and to evaluate defensive strategies that are applicable in real contexts.

\vspace{0.5cm}

\section{Modeling of Realistic Adversarial Attacks}
\label{sec:modeling}

We analyze adversarial attacks against ML-NIDS by combining and extending the taxonomies of Laskov et al.~\cite{Laskov:Practical} and Huang et al.~\cite{Huang:Adversarial}. To model the realistic capabilities of an attacker, we introduce the concept of ``power'' that denotes how much control the attacker has on the target detection system. 

We identify five elements on which the attacker has power:

\begin{itemize}
    \item \textit{Training Data.} It represents the ability to access the dataset used to train the ML-NIDS. It can come in the form of read, write, or no access at all.
    
    \item \textit{Feature Set.} It refers to the knowledge of the features analyzed by the ML-NIDS to perform its detection. It can come in the form of none, partial or full knowledge.
    
    \item \textit{Detection Model.} It describes the knowledge of the (trained) ML model integrated into the NIDS that is used to perform the detection. This knowledge may be none, partial or full.
    
    \item \textit{Oracle.} This element denotes the possibility of obtaining feedback from the output produced by the ML-NIDS to an attacker's input. This feedback can be limited, unlimited or absent.
    
    \item \textit{Manipulation Depth.} It describes the nature of the adversarial manipulation, that may modify the analyzed traffic (problem space) traffic level or one or more features (feature space.
\end{itemize}
We describe each element from the perspective of an attacker that aims at thwarting a ML-NIDS. The conclusions are summarized in Section~\ref{sec:considerations}.

We assume the typical ML-NIDS scenario of an attacker that has already established a foothold into one (or more) device of the monitored network perimeter, and intends to maintain (and expand) their access by remaining undetected through evasion attacks~\cite{Wu:Evading, Gardiner:Malware, Apruzzese:Evaluating, Martins:Adversarial}.

\subsection{Training data}
\label{sec:training}
Power on the training data comes in two different forms: \textit{read} access, which may allow an attacker to reproduce the training phase and obtain a similar detector to the one adopted by the organization; or \textit{write} access, which can be leveraged to perform poisoning attacks by either injecting new data, or modifying existing samples.

Obtaining power on the training data depends on the ML-NIDS solution adopted by the target organization: the detector can either be developed and managed entirely in-house, or can be acquired from a third-party vendor.

In the former case, the attacker may be theoretically able to retrieve the training data by identifying the device hosting the dataset, and then inspecting or exfiltrating the actual data. In practice, a similar machine is well protected through restrictive access policies~\cite{Chernikova:Adversarial}, and/or segregated in a dedicated network segment that is inaccessible by the hosts of the compromised network.
In the latter case, the ML component is trained on datasets of the vendor that are not accessible by the attacker unless they also compromise the vendor's network, which is unlikely. However, we stress that if attackers manage to seize control on the training dataset adopted by a third-party provider, then they could potentially launch adversarial attacks against any enterprise that is adopting the specific solution of the (compromised) provider. For this reason, providers of ML-based solutions for cybersecurity must protect their assets as much as possible.

A different circumstance occurs when the ML-NIDS is periodically retrained on new data. In this case, an attacker that has acquired a foothold into the organization may generate some malicious traffic that may affect the detection, thus granting some (indirect) write access to the training data. This attack strategy is unfeasible without direct read access to the training data. Attackers cannot be sure that the produced samples are truly used to retrain the ML-NIDS unless they can inspect the composition of the training dataset.

We can conclude that acquiring any form of reliable access (either \textit{read} or \textit{write}) to the training data used to train (or re-train) the ML-model is not very realistic~\cite{Shetty:Simulation, Li:Chronic, Garg:Obliviousness}.
Some power on the training data is obtainable only if the NIDS and corresponding training set are entirely managed by the target organization.

\subsection{Detection Model}
\label{sec:model}
With power on the detection model, an attacker can determine the internal configuration of the ML method, alongside the decision boundaries learnt after its training phase. White-box attacks denote scenarios where such knowledge is available to the adversary. 
The detection model represents the core component of the ML-NIDS, which is deployed on machines whose access requires high administrative privileges, and which can be contacted only by few selected devices~\cite{Khalil:Optimal}. 
Hence, it is unrealistic to assume that an infected host can allow the attacker to access the ML-NIDS containing its detection model~\cite{Shetty:Simulation, Yang:Adversarial}. 

An attacker could perform reconnaissance and lateral movements~\cite{Apruzzese:Pivoting} to understand the network topology and eventually obtaining access to the NIDS machine. 
However, even in this unlikely case, to acquire power on the detection model the attacker must be able to inspect the underlying source-code. If the ML-NIDS is a commercial product, such code may not be observable; whereas if the ML-NIDS is developed in-house, then the (human readable) source-code may be located on a different machine~\cite{Shetty:Simulation,Chernikova:Adversarial}.

We exclude the possibility of directly modifying the target ML-NIDS: in a similar circumstance, the entire defensive system would be completely overthrown and at complete disposal of the attacker. Any rational attacker who already managed to take control over the detection system will be in a position to achieve all of its goals in a more reliable and direct way rather than through subtle adversarial attacks.

In summary, it is unlikely that an adversary may get power on the internal configuration of the ML model (which is the case of white-box attacks).

\subsection{Feature Set}
\label{sec:feature}
By knowing the features that represent a given sample, the attacker is able to determine which operations are required to generate an adversarial example. For instance, if the ML-NIDS analyzes the duration of network communications, then an attacker can alter the length of the sessions between the controlled hosts.
Obtaining power on the actual set of features used by the ML model faces similar challenges of gaining access to the trained detection model~\cite{Shetty:Simulation, Yang:Adversarial, Chernikova:Adversarial}. 
A similar scenario occurs because the actual features representing each sample and used to perform the detection are specified only at inference time.

However, attackers may leverage their domain expertise to guess which features are likely to be analyzed by the detector~\cite{Han:Practical, Chernikova:Adversarial}. Although each ML-NIDS is unique, they all analyze network traffic either in the form of raw network packets or as derived network metadata. Hence a ML-NIDS may employ feature sets with many similarities and overlaps~\cite{Apruzzese:Evaluating, Wu:Evading, Shanthi:Detection, Lashkari:Characterization}.
Therefore, it is reasonable to assume that expert attackers will leverage such intelligence to estimate with high probability some features utilized by the actual detector, which will impact the performance. The adversarial samples will then revolve around perturbations of these features. 

It is important to establish the relationship between power on the training data (see Section~\ref{sec:training}), and power on the feature set. The former does not necessarily lead to power on the latter. Manipulating the training data with perturbations of existing samples causes the modification of their features with consequences on the training phase. However, our definition of ``power on the feature set'' in Section~\ref{sec:modeling} denotes the \textit{knowledge} about the feature set. As an example, if attackers have write access they can inject (directly or indirectly) new samples to the training data, but they would not know the actual features considered by the ML detector. Hence, write access to the training dataset does not lead to power on the feature set.

The situation may differ when an attacker has a read access on the training dataset, and the features are distinguishable. In such a case, an attacker may acquire some knowledge and power on the feature set. However, it is possible that the actual features used by the model are computed at inference time~\cite{Apruzzese:Addressing}. For example, an organization may store the training dataset in the form of raw packet captures (PCAP), but the detector may be trained on network flows that are generated right before the training (or testing) phase. In such a scenario, even if attackers have power on the training data, they would not know the complete feature set used for the detection process although they can guess it and determine a portion of the actual feature set.

In summary, it is realistic to assume an attacker that has some power on the feature-set used by the ML-NIDS; having complete knowledge on this element is a tough challenge. Finally, it is unlikely that an attacker is completely oblivious of the composition of the feature-set.

\subsection{Oracle}
\label{sec:oracle}
An attacker that has oracle power is able to reverse-engineer the target ML model by submitting some inputs and observing the corresponding output (see Section~\ref{sec:attacks}). In the specific case of ML-NIDS, to obtain oracle power the attacker faces two obstacles: not triggering other detection mechanisms, and extracting meaningful information from the input/output feedback.

\textbf{Obstacle 1: Avoiding Detection.}
Modern organizations protect their networks through multiple defensive layers~\cite{Pierazzi:Online}. Hence, an attacker must operate in such a way to avoid being detected by these additional detection schemes. To this purpose, the attacker should aim at minimizing the amount of queries issued to the target NIDS~\cite{Kuppa:Black, Ilyas:Black}, since each query requires to create and send some additional anomalous traffic to the target network. 
Furthermore, these queries should be performed in a low-and-slow approach because excessive queries in a short time frame may easily trigger alerts by detection systems that leverage simple statistical approaches to model the normal network traffic~\cite{Pierazzi:Online, More:Semantic}.
Similar methodologies require an extended amount of time, up to days or weeks. These operations increase the probability that the attacker is detected. At the very least, they increase the length and cost of the offensive campaign~\cite{Kang:Spiffy}. 
In summary, an attacker can interact with the NIDS by sending some queries, but any rational attacker will try its best to limit the number of interactions.

\textbf{Obstacle 2: Acquiring Feedback.}
The output of the detection may not be directly observable by an attacker. When a NIDS identifies a malicious event, it generates \textit{alerts} which are only notified to security administrator~\cite{Liao:Review} through the NIDS console. In other words, even if attackers perturb some input samples, they cannot reliably receive any feedback from the NIDS. To witness the results of their own actions, attackers need to wait until the human operator, after triaging the alerts raised by the malicious samples, applies policies that specifically address the samples issued by the attacker. Thus, in these circumstances the adversary must rely on the defender's reactions, which is an unreliable attacking approach that may require long timespans.
We identify three situations where the attacker can directly observe some information about the classification output generated by the NIDS after a given query.
\begin{itemize}
   
    \item The first scenario requires the NIDS to be integrated in some reactive defensive mechanism that is able to automatically stop the detected malicious traffic. 
    In this case an attacker can obtain the input/output pairing, for example by investigating which network communication of the controlled machines are (or not) received by the external hosts. Although the use of similar defensive systems is common in modern environments, we observe that the attacker must be aware that these mechanisms operate with the NIDS and that they are configured to block the specific traffic generated by the attacker. Acquiring such information requires significant intelligence expertise.
    
    \item The second scenario requires the attacker to gain access to the NIDS logs or its console, but this is unlikely because such data are accessible only by the NIDS administrator.
    
    \item The third scenario requires the organization to rely on a commercial NIDS. In a similar case, the attackers could acquire the same NIDS and deploy it in a controlled environment where they have complete freedom: they may still be unable to determine the inner configurations of the ML components, but they would not be subject to any limitations of the query. A similar scenario is rather unrealistic as the attacker needs to know the NIDS product used by the target organization, which may require extended intelligence operations. They also must acquire the product by the third party vendor, which increases the cost of the offensive campaign; finally, they have to reproduce a network environment that resembles that of the target organization. All these requirements make this scenario feasible only for skilled and highly motivated adversaries.
\end{itemize}

We observe that attackers willing to acquire oracle power must overcome both obstacles (that is, avoiding detection, and receiving feedback). Attackers that can perform an unlimited amount of queries without being detected do not have any oracle power if they cannot reliably determine the input-output association. Similarly, attackers that can obtain the input-output association but may be detected during the process.

We can conclude that using the NIDS as an oracle is a tough challenge. In real scenarios, motivated attackers may get some feedback but with many limitations (e.g., few queries, or uncertain ML predictions), while a complete oracle power is feasible only if the organization employs a commercial NIDS, or if the NIDS itself is compromised.

\subsection{Manipulation depth}
\label{sec:space}

The last element on which the attacker has power involves the data manipulation capabilities: the (adversarial) perturbations can be introduced either at the raw traffic level, or after the network data is being transformed into its higher-level feature representation.
Indeed, an emerging topic in adversarial ML literature is the differentiation between \textit{feature-}space and \textit{problem-}space attacks~\cite{Pierazzi:Intriguing, Ibitoye:Threat, Inkawhich:Feature}.
The former implies that the adversarial perturbation is applied directly to the sample that is provided as an input to the ML model. Problem-space attacks involve obtaining perturbations by performing all the operations at the lower data level. As a practical example, consider a ML-NIDS that works on network flows: a feature space attack may involve increasing the value of one feature of the flow sample (e.g., the duration). A problem-space attack requires modifying the malware logic so as to produce network packets that, when exported to network flows, result in samples with increased duration (e.g., by adding some communication delays).
In summary, feature-space attacks are a higher-level abstraction of the attacker's workflow focusing only on its intended results, whereas problem-space attacks involve the reproduction of the entire malicious procedure.

For these reasons, attacks at the problem-space are more representative of a realistic scenario. The only way in which an adversary could perform a real feature-space attack is by manipulating the conversion of the raw traffic data into its feature representation, which requires full power on the ML component.

\subsection{Final considerations}
\label{sec:considerations}

\begin{figure}[!htbp]
    \centering
    \includegraphics[width=1\columnwidth]{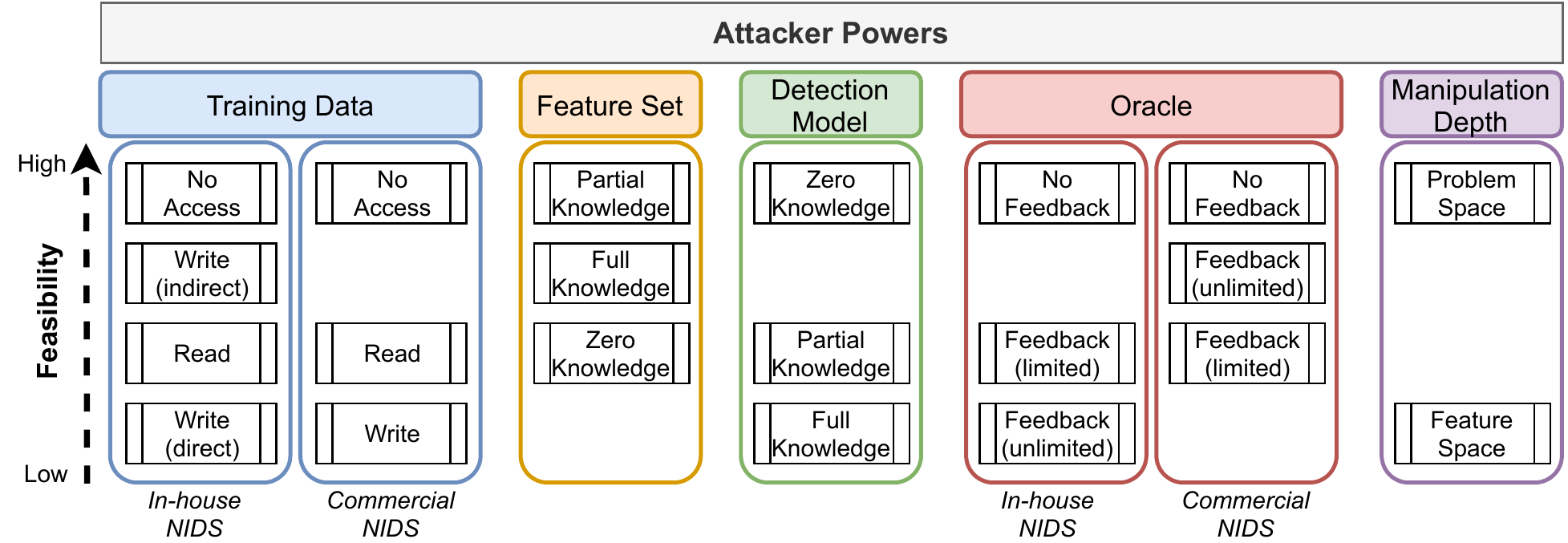}
    \caption{Feasibility of each Power available to the attacker.}
    \label{fig:power_detail}
\end{figure}

By taking into account the attacker power on all five elements, we summarize the considerations in Figure~\ref{fig:power_detail} representing the powers on the target NIDS. For each power, the figure shows the feasibility of its different forms as in  Section~\ref{sec:modeling}. These forms are vertically ordered on the basis of their realistic feasibility: the most likely are placed in higher positions, and the least likely in lower positions. For the Training Data and Oracle powers, we distinguish the commercial NIDS on the right, and the in-house solutions on the left.

We remark that any attack is possible, but some require a higher amount of resources to be carried out, which may discourage their actuation. The aim of Figure~\ref{fig:power_detail} is to help determining which circumstances are more likely to occur in a real adversarial attacks against ML-NIDS.

In the most realistic attacks the attacker manipulates the real network traffic (where the perturbations occur at the \textit{problem space}, rather than at the \textit{feature space}) with just a \textit{partial knowledge} of the features, \textit{no feedback} from the NIDS, and \textit{no access} to the training data. These powers do not require an attacker to previously compromise the system nor to acquire knowledge about some proprietary and possibly well protected information about the NIDS and the target organization. Hence, similar attacks can be deployed even by attackers with an average skill and at low costs.

Adversarial attacks relying on the oracle power are feasible only if the target organization leverages a commercial NIDS that the attacker can also buy to freely experiment with. However, even in this situation the feasibility might be limited by non-standard configurations employed in the target organization that the attacker will likely not know and replicate on their testing environment.

Attacks that require power on the training data or on the detection model of a NIDS that is built and maintained in-house by the target organization, or a full knowledge of the feature set, are unrealistic, since these powers can only be obtained by an attacker that already managed to compromise the systems belonging to the target network that store these information.
The same considerations also apply to attacks that require the use of the detection system as an oracle. In principle, a similar result can be achieved by an attacker that submits malicious traffic samples to a network intrusion prevention system (i.e., a detection system that is also configured to block malicious network communications). However this allows only to achieve partial knowledge, increases the cost and duration of the attack campaign and also increases the likelihood that these ancillary attack activities might trigger some other alert. 

The most unrealistic attacks are those requiring full power over the training data or on the detection model of a commercial IDS, since the attacker should be able to compromise the vendor of the commercial defensive solution. Even attacks that require power on the detection model of a commercial NIDS are quite unrealistic, because they require an attacker to arbitrarily modify the ML-based detector executed by the commercial NIDS appliance acquired by the target organization. Finally, attacks that can only be performed at the feature space fall in this category, since the attacker would need to compromise the component of the detector that extracts features from the raw traffic.

We conclude with an important observation: \textit{poisoning} attacks at the training-phase can only be performed if the attacker has power (in the form of write access) on the training data. Otherwise, the attacker is limited to attacks at inference-time. 

Poisoning attacks can be extremely powerful, but they can be avoided by negating attacker power on the training set. On the other hand, devising threat models for attacks at inference-time is more complex because the attacker can leverage a wider array of powers. Hence, attacks at inference time may be less disruptive, but they are more difficult to prevent.
\section{Realistic Evaluation of Existing Attacks}
\label{sec:analysis}

We now evaluate the adversarial attacks against NIDS proposed in the literature by applying the proposed modeling guidelines, with the goal of analyzing the maturity and realism of existing examples. We initially describe some operations required to simulate realistic adversarial attacks. Then we provide an overview of the state of the art (Section~\ref{sec:overview}), and finally describe the details of some cases (Section~\ref{sec:case_studies}).

In order to reproduce realistic adversarial attacks against NIDS it is necessary to ensure that the perturbations maintain the malicious logic of the original sample. Indeed, to evade a ML detector it would suffice to generate a completely new malware variant that is not represented by any malicious sample contained in the training dataset.
However, doing so would defeat the purpose of adversarial attacks, which involves the application of small and imperceptible modifications~\cite{Su:One, Papernot:SoK}. Hence, the modified samples need to only slightly differ from their original variants, and they must also maintain their underlying malicious logic without triggering other detection mechanisms~\cite{Piplai:Nattack, Wu:Evading, Apruzzese:Evaluating, Kuppa:Black, Usama:Generative}.
While such conditions are verifiable for attacks performed in the problem-space, attacks performed at the feature-space require additional verification steps~\cite{Han:Practical}. 

Furthermore, when simulating attacks at the feature-space it is also needed to check that all inter-dependencies between features are maintained, and that the feature values of the perturbed samples do not result in impossible numbers: for example, the size of a TCP packet cannot exceed 64KBytes, while some network flow collectors have fixed thresholds for the maximum flow duration~\cite{Peng:Adversarial, Apruzzese:Pivoting}.

In order to evaluate realistic adversarial attacks against NIDS, it is necessary to ensure that all these properties are preserved~\cite{Hashemi:Towards, Lin:IDSGAN}.

\subsection{Overview}
\label{sec:overview}

We perform an extensive literature survey on adversarial attacks against ML-NIDS. To the best of our knowledge, the overall results reported in Table~\ref{tab:evaluation_table} represent the state-of-the-art until March 2021.

\begin{table*}[!htbp]
    \centering
    \caption{Characteristics of existing Adversarial Attacks against ML-NIDS.}
    \label{tab:evaluation_table}
    \resizebox{1\columnwidth}{!}{
        \begin{tabular}{|c|c?c|c|c|c|c?c|c|c|}
             \hline
             \multirow{3}{*}{Paper}  & \multirow{3}{*}{Year} & \multicolumn{5}{c?}{\textbf{Power}} & 
             \multirow{3}{*}{\begin{tabular}{@{}c@{}} Constraints \\ Verified? \end{tabular}} &
             
             \multirow{3}{*}{Type} & \multirow{3}{*}{Dataset} \\ \cline{3-7}

             & & \begin{tabular}{@{}c@{}} \textit{Training} \\ \textit{Data} \end{tabular} &  
             \textit{Feature set} & 
             \begin{tabular}{@{}c@{}} \textit{Detection} \\ \textit{model}\end{tabular} & 
             \textit{Oracle} & 
             \begin{tabular}{@{}c@{}} \textit{Manipulation} \\ \textit{depth} \end{tabular} & & &
             \\ \hline \hline

            Kloft et al.~\cite{Kloft:Online} & \pdate{2010} & \power{W} & \power{\NA} & \power{Full} & \power{\NA} & \power{feature} & \xmark & poisoning & \dataset{KDD99}  \\ \hline
            Biggio et al.~\cite{Biggio:Security2} & \pdate{2013} & \power{\NA} & \power{Full} & \power{Partial} & \power{\NA} & \power{feature} & \xmark & inference & \dataset{custom} \\ \hline
            Sethi et al.~\cite{Sethi:Data} & \pdate{2018} & \power{\NA} & \power{Full} & \power{\NA} & \power{\NA} & \power{feature} & \cmark & inference & \dataset{KDD99} \\ \hline
            Warzynski et al.~\cite{Warzynski:Intrusion} & \pdate{2018} & \power{\NA} & \power{Full} & \power{Full} & \power{\NA} & \power{feature} & \power{\xmark} & inference & \dataset{KDD99} \\ \hline
            Apruzzese et al.~\cite{Apruzzese:Evading, Apruzzese:Evaluating} & \pdate{2018} & \power{\NA} & \power{Partial} & \power{\NA} & \power{\NA} & \power{feature} & \cmark & inference & \begin{tabular}{@{}c@{}} \footnotesize{CTU13, Botnet2014} \\ \footnotesize{IDS2017, CICIDS2018} \end{tabular} \\ \hline
            Lin et al.~\cite{Lin:IDSGAN} & \pdate{2018} & \power{\NA} & \power{\NA} & \power{\NA} & \power{\Large{$\infty$}} & \power{feature} & \cmark & inference & \dataset{KDD99} \\ \hline
            Li et al.~\cite{Li:Chronic} & \pdate{2018} & \power{R, W} & \power{Partial} & \power{\NA} & \power{\NA} & \power{feature} & \xmark & poisoning & \dataset{KDD99, Kyoto\cite{Song:Statistical}} \\ \hline
            Wang et al.~\cite{Wang:Deep} & \pdate{2018} & \power{\NA} & \power{Full} & \power{Full} & \power{\NA} & \power{feature} & \xmark & inference & \dataset{KDD99} \\ \hline
            Yang et al.~\cite{Yang:Adversarial} & \pdate{2018} & \power{R} & \power{Partial} & \power{\NA} & \power{\Large{$\infty$}} & \power{feature} & \cmark & inference & \dataset{KDD99} \\ \hline
            Marino et al.~\cite{Marino:Adversarial} & \pdate{2018} & \power{\NA} & \power{Full} & \power{Full} & \power{\NA} & \power{feature} & \power{\xmark} & inference & \dataset{KDD99} \\ \hline
            Apruzzese et al.~\cite{Apruzzese:Addressing} & \pdate{2019} & \power{R, W} & \power{Partial} & \power{\NA} & \power{\NA} & \power{feature} & \xmark & poisoning & \dataset{CTU13} \\ \hline
            Hashemi et al.~\cite{Hashemi:Towards, hashemi2020enhancing} & \pdate{2019} & \power{\NA} & \power{Full} & \power{Full} & \power{\NA} & \power{feature} & \xmark & inference & \dataset{IDS2017} \\ \hline

            Usama et al.~\cite{Usama:Generative} & \pdate{2019} & \power{\NA} & \power{\NA} & \power{\NA} & \power{\Large{$\infty$}} & \power{feature} & \cmark & inference & \dataset{KDD99} \\ \hline

            Khamis et al.~\cite{Khamis:Investigating} & \pdate{2019} & \power{\NA} & \power{Full} & \power{Full} & \power{\NA} & \power{feature} & \xmark & inference & \dataset{UNSW-NB15} \\ \hline
            Clemens et al.~\cite{Clements:Rallying} & \pdate{2019} & \power{\NA} & \power{Full} & \power{Full} & \power{\NA} & \power{feature} & \xmark & inference & \dataset{Kitsune} \\ \hline
            
            Ibitoye et al.~\cite{Ibitoye:Analyzing} & \pdate{2019} & \power{\NA} & \power{Full} & \power{Full} & \power{\NA} & \power{feature} & \xmark & inference & \dataset{BoT-IoT\cite{Koroniotis:Towards}} \\ \hline
            Aiken et al.~\cite{Aiken:Investigating} & \pdate{2019} & \power{\NA} & \power{Partial} & \power{\NA} & \power{\NA} & \power{problem} & \cmark & inference & \dataset{IDS2017} \\ \hline
            Yan et al.~\cite{Yan:Automatically} & \pdate{2019} & \power{\NA} & \power{\NA} & \power{\NA} & \power{\Large{$\infty$}} & \power{feature} & \cmark & inference & \dataset{KDD99, IDS2017} \\ \hline
            Martins et al.~\cite{Martins:Analyzing} & \pdate{2019} & \power{\NA} & \power{Full} & \power{Full} & \power{\NA} & \power{feature} & \xmark & inference & \dataset{KDD99, IDS2017} \\ \hline
            Usama et al.~\cite{Usama:Black} & \pdate{2019} & \power{\NA} & \power{\NA} & \power{\NA} & \power{\Large{$\infty$}} & \power{problem} & \cmark & inference & \dataset{Tor-nonTor\cite{Lashkari:Characterization}} \\ \hline
            Peng et al.~\cite{Peng:Adversarial} & \pdate{2019} & \power{\NA} & \power{\NA} & \power{\NA} & \power{\Large{$\infty$}} & \power{feature} & \cmark & inference & \dataset{KDD99, IDS2017} \\ \hline
            Kuppa et al.~\cite{Kuppa:Black} & \pdate{2019} & \power{\NA} & \power{Partial} & \power{\NA} & \power{100s} & \power{feature} & \cmark & inference & \dataset{CICIDS2018} \\ \hline
            Wu et al.~\cite{Wu:Evading} & \pdate{2019} & \power{\NA} & \power{Partial} & \power{\NA} & \power{10s} & \power{problem} & \cmark & inference & \dataset{CTU13} \\ \hline
            Ayub et al.~\cite{Ayub:Model} & \pdate{2020} & \power{\NA} & \power{Full} & \power{Full} & \power{\NA} & \power{feature} & \xmark & inference & \dataset{IDS2017, TRAbID2017\cite{Viegas:Toward}} \\ \hline
            Novo et al.~\cite{Novo:Flow} & \pdate{2020} & \power{\NA} & \power{Full} & \power{Full} & \power{\NA} & \power{feature} & \cmark & inference & \dataset{MTA\cite{MTA:URL}} \\ \hline
            Chernikova et al.~\cite{Chernikova:Adversarial} & \pdate{2020} & \power{\NA} & \power{Partial} & \power{\NA} & \power{\Large{$\infty$}} & \power{feature} & \cmark & inference & \dataset{CTU13} \\ \hline
            Sadeghzadeh et al.~\cite{Sadeghzadeh:Adversarial} & \pdate{2020} & \power{\NA} & \power{Full} & \power{Full} & \power{\NA} & \power{problem} & \cmark & inference & \dataset{ICSX2016} \\ \hline
            Chernikova et al.~\cite{Chernikova:Adversarial} & \pdate{2020} & \power{\NA} & \power{Full} & \power{Full} & \power{\NA} & \power{feature} & \cmark & inference & \dataset{CTU13} \\ \hline
            Piplai et al.~\cite{Piplai:Nattack} & \pdate{2020} & \power{\NA} & \power{Full} & \power{Full} & \power{\NA} & \power{feature} & \xmark & inference & \dataset{BigDataCup2019\cite{Janusz:IEEEBigDataCup}} \\ \hline
            Alhajjar et al.~\cite{Alhajjar:Adversarial} & \pdate{2020} & \power{\NA} & \power{Full} & \power{Full} & \power{\NA} & \power{feature} & \xmark & inference & \dataset{KDD99, UNSW-NB15} \\ \hline
            Apruzzese et al.~\cite{apruzzese2020deep} & \pdate{2020} & \power{\NA} & \power{Partial} & \power{None} & \power{20s} & \power{feature} & \xmark & inference & \dataset{CTU13, Botnet2014} \\ \hline
            Han et al.~\cite{Han:Practical} & \pdate{2020} & \power{\NA} & \power{Partial} & \power{\NA} & \power{100s} & \power{problem} & \cmark & inference & \dataset{Kitsune, IDS2017} \\ \hline
            
            Pawlicki et al.~\cite{pawlicki2020defending} & \pdate{2020} & \power{\NA} & \power{Full} & \power{Full} & \power{\NA} & \power{feature} & \xmark & inference & \dataset{IDS2017} \\ \hline
            
            Shu et al.~\cite{shu2020generative} & \pdate{2020} & \power{\NA} & \power{Full} & \power{Partial} & \power{50s} & \power{feature} & \xmark & inference & \dataset{IDS2017} \\ \hline
            
            Papadopoulos et al.~\cite{papadopoulos2021launching} & \pdate{2021} & \power{R, W} & \power{Full} & \power{\NA} & \power{\NA} & \power{feature} & \xmark & poisoning & \dataset{Bot-IoT} \\ \hline
            
            Pacheco et al.~\cite{pacheco2021adversarial} & \pdate{2021} & \power{\NA} & \power{Full} & \power{Full} & \power{\NA} & \power{feature} & \xmark & inference & \dataset{UNSW-NB15} \\ \hline
            
            Anthi et al.~\cite{Anthi:Adversarial} & \pdate{2021} & \power{R} & \power{Full} & \power{\NA} & \power{\NA} & \power{feature} & \xmark & inference & \dataset{ICSX2016} \\ \hline
            
            Papadopoulos et al.~\cite{papadopoulos2021launching} & \pdate{2021} & \power{\NA} & \power{Full} & \power{Full} & \power{\NA} & \power{feature} & \xmark & inference & \dataset{Bot-IoT} \\ \hline
            
        \end{tabular}
    }

\end{table*}

For each paper, we report its publication date, and the power of the considered attacker for each element described in Section~\ref{sec:modeling}.
\begin{itemize}
    \item \textit{Training Data}: the attacker can have Read, Write or no access (R, W, or ``\NA'', respectively).
    \item \textit{Feature Set} and \textit{Detection Model}: The attacker can have Full, Partial or no (denoted with ``\NA'') knowledge.
    \item \textit{Oracle}: papers where the NIDS is used as an oracle always assume that the attacker has full feedback on the the input-output association. We use ``$\infty$'' to denote papers that do not set any boundary on the amount of queries available to the attacker; for proposals that consider a constrained attacker, we report the order of magnitude of the required interactions (e.g., 10s denotes tens of queries); we use ``\NA'' for papers where the attacker does not use the NIDS as an oracle.
    \item We also report if the manipulation occurs at the \textit{feature} or \textit{problem} space.
\end{itemize}

Moreover, we specify whether the authors consider perturbations that preserve the maliciousness and integrity of the original sample ({\cmark}) or not ({\xmark}). Finally, in the two rightmost columns we report the type of attack (inference or poisoning), and the data sets used for the experimental testbed. If a paper considers different attack scenarios, it will have more entries in the table.
As an example, let us consider the scenario proposed in~\cite{Li:Chronic}: this paper presents a poisoning attack where the adversary is assumed to have both Read and Write access to the training data, and has partial knowledge of the features used by the analysis model; however, they have no knowledge on the internal configuration of the detector, and cannot use it as an oracle. Finally, the proposed attack involves manipulations occurring in the feature space and the authors do not verify that the adversarial samples preserve their realistic integrity.

From this table we can express several considerations.
The initial studies considered an outdated and widely deprecated dataset (KDD99~\cite{Sharafaldin:Toward}). More recent works include additional datasets in their experiments, such as the CTU-13~\cite{Garcia:CTU}, the Kitsune~\cite{Clements:Rallying}, the UNSW-NB15~\cite{UNSWNB15:Dataset}, the ICSX2016~\cite{ICSXVPN:Dataset}, or the very recent IDS2017 and CICIDS2018~\cite{Sharafaldin:Toward}. This is an important improvement, because evaluating adversarial attacks on multiple datasets increases the impact and value of the experiments. Furthermore, by considering more recent data, these results allow the researchers to understand the effectiveness of their threats in modern defensive scenarios.

We also observe that just a minority of proposals considers poisoning attempts~\cite{Kloft:Online, Li:Chronic, Apruzzese:Addressing}, while the majority of them focuses on evasion attacks at inference time. This is a relevant trend because, as discussed in Section~\ref{sec:training}, poisoning attacks are difficult to perform in real scenarios due to the difficulty faced by attackers to acquire some power on the training data.

Assuming that the ML-NIDS can be used as an oracle that answers to an unlimited amount of queries is possible only if an attacker can acquire a perfect replica the detector. However, only some recent papers (e.g.,~\cite{Han:Practical, Wu:Evading, Kuppa:Black}) consider attackers with a limited number of queries. This is an important step towards realistic security scenarios.

A large amount of proposals verify whether the modified samples preserve or not their integrity and their malicious logic. Few papers regard attacks at problem space (e.g.,~\cite{Usama:Black,Sadeghzadeh:Adversarial, Han:Practical}). This choice evidences the possibility of novel research opportunities that can evaluate more complex but also more realistic adversarial samples.

\subsection{Case Studies}
\label{sec:case_studies}

We consider three papers~\cite{Aiken:Investigating, Wu:Evading, Li:Chronic} from Table~\ref{tab:evaluation_table} because they represent meaningful examples for adversarial attacks at different degrees of feasibility. Then, for comparison purposes we analyze a famous case of a real adversarial attack against a real detector~\cite{liang2016cracking}.

The work in~\cite{Aiken:Investigating} involves evasion attacks at inference time against NIDS. The adversary has very little power on the target system: they have no access to the training set, and cannot interact with the detector in any way (neither to inspect its internal configuration, nor to leverage it as an oracle). The only assumption is that the attacker knows some of the features used by the detection mechanism. Thus, the strategy consists in modifying the payloads of the raw network traffic (resulting in a problem-space attack), so as to induce perturbations of these features that induce the model to misclassify the malicious network traffic. In their experiments, Aiken et al.~\cite{Aiken:Investigating} show the performance drop of state-of-the-art classifiers against these attacks: the baseline $99.9\%$ detection rate decreases to an unacceptable $70\%$.
There are several elements that make the scenario described in this paper as very realistic. Firstly, the attacker is not assumed to have access to some of the most protected devices in the target system (that is, the detector, and the server hosting the training data); the only power available to the attacker is the knowledge of a subset of the features adopted by the classification mechanism, which is a feasible assumption (see Section~\ref{sec:feature}). Furthermore, the operations performed to apply the perturbations are easily achievable for any attacker that has established a foothold in a network environment. Finally, the target detector is trained on a recent dataset (the IDS2017), representing modern network environments.
We conclude that the attack represented in this paper portrays a complete and realistic adversarial scenario.

The authors of~\cite{Wu:Evading} consider attacks at inference time against botnet detectors. The adversary plans to use an autonomous deep reinforcement learning agent to evade a classifier based on network flows. The paper assumes an attacker that is able to inspect the feedback of the detector to a given sample (they can use it as an oracle) with no limitation to the amount of queries that can be performed. The attacker also guesses that the detector analyzes network flows, and they are aware of the possible features employed. However, they do not have any power on the training data, and have no knowledge of the ML-model integrated in the NIDS. The samples are generated at the problem space, because the agent modifies the raw (malicious) network packets by adding redundant data in the payload, therefore altering the network flows that are forwarded to the NIDS. With these settings, the authors of~\cite{Wu:Evading} show that the proposed agent is able to generate adversarial samples that evade detection in almost $80\%$ of the cases, by requiring only dozens of queries.
Based on these assumptions, we consider this paper to represent a realistic but less feasible scenario than the one in~\cite{Aiken:Investigating}, due to the additional presence of the oracle power. As explained in Section~\ref{sec:oracle}, receiving direct feedback from the NIDS requires that the attacker either (i) acquires the same NIDS and uses it in a dedicated environment (if the NIDS used by the target is produced by a third-party vendor); or (ii) that the NIDS also integrates an IPS that can be leveraged to determine which communications are blocked. In the former case, the attacker needs to invest resources to acquire and deploy the NIDS, which is feasible but increases the cost of the campaign. In the latter case, the attacker must ensure that the performed queries do not trigger an excessive amount of alarms: we consider the few dozens queries required in~\cite{Wu:Evading} to be an acceptable number that is unlikely to induce security personnel to manually investigate the infected devices and triage them.

The third case study~\cite{Li:Chronic} describes a poisoning attack where the attackers are assumed to have power on training data. They know the entire composition of the data used to train the model because they have full reading access to the training dataset. They also have write access. Although they cannot change the label (as in label flipping~\cite{taheri2020defending}) or modify the features of existing training samples, they are allowed to inject new samples in the training dataset. The adversary also knows the features used by the detector - which correspond to the features representing the samples of the training dataset. The attacker thus plans to add some adversarial samples in the training set in order to modify the decision boundaries of the detection model, thus ensuring that malicious samples remain undetected. By executing these operations, Li et al.~\cite{Li:Chronic} show that the accuracy of detectors could drop from over $95\%$ to nearly $50\%$, making the target detector impractical.
We consider this scenario as unrealistic. As explained in Section~\ref{sec:training}, obtaining both read and write access to the training data is a daunting task even for expert adversaries. Having complete knowledge of the features used by the detection model is not realistic: even if the attacker has access to the training data and knows the features associated to each sample, it is likely that the feature-set used by the detection model presents some differences. For example, it may include some derived features computed right before the training operations (\cite{Apruzzese:Addressing}), or it may even discard some features that lead to unfavorable performance. Finally, we remark that the targeted detectors are trained on two outdated datasets (KDD99 and Kyoto2006) that do not capture the characteristics of the network traffic generated by modern organizations and recent attacks~\cite{Sharafaldin:Toward}.

Finally, we consider an interesting attack against a real detection system that is embedded into the phishing detector of the Google Chrome browser~\cite{liang2016cracking}. Although this work is not included in Table~\ref{tab:evaluation_table} because the target system is not a NIDS, we find it useful to summarize its circumstances because it can be inspirational for future researches.
The considered scenario involves a ML-detection system that is deployed on a popular browser that is obtainable by any user. The (trained) detection model is embedded into the application, so it is impossible to affect the training phase. Furthermore, the underlying application code is not readable, hence it is impossible to determine which features are used to perform the analysis, nor to acquire any information about the actual detection model. However, the accessibility of the application, and therefore of its embedded detection system, allows a user to have complete \textit{oracle} power, with an unlimited amount of queries and direct access to the feedback of the input/output pair. In other words, a user can craft a Web page, and then visit such page with the browser: if the browser alerts the user, then it means that the page is considered as malicious; otherwise, the page  is considered as benign.
An attacker with a similar power can create a wide array of pages to reverse engineer the classifier used to perform the detection. Consequently, the authors of~\cite{liang2016cracking} were able to gain important information about the detection system, such as which ML algorithm was used, the parameters, and the most significant classification features. This allowed the attackers to identify the criteria used to pinpoint whether a Web page was malicious or not and it allowed to determine how to evade a similar system. The authors were able to camouflage a Web page (in the \textit{problem} space) so that its malicious score dropped from 0.99 to 0.45.
The attack in~\cite{liang2016cracking} is an example of a real use case where the attackers have no power on the training data, the feature set, and the detection model; they operate at the problem space, but they have complete oracle power. The vulnerability is represented by the deployment of the detection system at the client level: a similar attack would be more difficult if the ML model was deployed on a dedicated, controlled and remote server. This is the typical case of a NIDS and motivates the difficulty of obtaining full oracle power in real circumstances.
\section{Conclusions}
\label{sec:conclusions}

Adversarial attacks represent a threat affecting the reliability of any cyber defense relying on artificial intelligence. Literature has shown the poor performance of ML-NIDS against adversarial perturbations, but few papers analyze this emerging menace by taking into account the realistic characteristics of modern environments. The consequence is that many threat models are practically unfeasible.
By assuming a cybersecurity perspective, we identify and model the different elements of the target systems that can be leveraged by an attacker to carry out an adversarial attack against ML-NIDS. We discuss each of these elements by describing the realistic circumstances that allow attackers to seize its control. The proposed model is then applied to analyze several papers presenting adversarial attacks against ML-NIDS. On the base of the identified parameters, we can easily identify papers considering scenarios that are more likely (or unlikely) to result in real adversarial attacks.
We can conclude that many papers assume adversarial threat models that are inapplicable to realistic ML-NIDS, but recent works consider some real obstacles that attackers need to overcome to bypass or perturb detection. This paper can guide researchers in devising threat models that are more representative of real defensive settings, and motivates the need for additional and more realistic research on adversarial attacks against ML-NIDS. The identification of the main defensive vulnerabilities and the prioritization against known adversarial attacks allows security experts to harden ML-based defensive systems. 
Our paper is specifically oriented to Network Intrusion Detection problems, but some analyses and conclusions can be applied to other cyber detection problems, such as phishing and malware detection. These contexts can represent interesting applications for extensions in future work.


\end{document}